\documentclass[aps,prb,twocolumn,longbibliography,groupedaddress,showpacs,floatfix,superscriptaddress,a4paper]{revtex4-2}

\usepackage[dvipsnames]{xcolor}
\definecolor{darkerBlue}{RGB}{30, 81, 149}
\usepackage[
    plainpages=false,
    pdfpagelabels,
    colorlinks=true,
    linkcolor=BrickRed,
    urlcolor=darkerBlue,
    citecolor=darkerBlue,
    pdftitle={The thermopower properties of interacting systems},
    pdfauthor={},
    pdfdisplaydoctitle=true,
    pdfduplex=DuplexFlipLongEdge
]{hyperref}

\usepackage{amsmath,amssymb}
\usepackage{bm}
\usepackage{braket}
\usepackage{orcidlink}
\usepackage{graphicx}
\usepackage{upgreek}
\usepackage{soul}

\begin{document}

\title{The thermopower properties of interacting systems}

\author{M. A. Habitzreuter \orcidlink{0000-0002-3534-1409}}
\affiliation{International Institute of Physics, Universidade Federal do Rio Grande do Norte, Natal, RN 59078-970, Brazil}

\author{Willdauany C. de Freitas da Silva \orcidlink{0000-0002-4133-2234}}
\affiliation{Departamento de F\'isica, Universidade Federal da Para\'iba, Joao Pessoa, PB 58051-900, Brazil.}

\author{Rodrigo A. Fontenele \orcidlink{0000-0002-8251-0527}}
\affiliation{Instituto de F\'isica, Universidade Federal do Rio de Janeiro, Rio de Janeiro, RJ 21941-972, Brazil}

\author{Natanael C. Costa \orcidlink{0000-0003-4285-4672}}
\email{natanael@if.ufrj.br}
\affiliation{Instituto de F\'isica, Universidade Federal do Rio de Janeiro, Rio de Janeiro, RJ 21941-972, Brazil}

\author{Thereza Paiva \orcidlink{0000-0002-4199-3809}}
\email{tclp@if.ufrj.br}
\affiliation{Instituto de F\'isica, Universidade Federal do Rio de Janeiro, Rio de Janeiro, RJ 21941-972, Brazil}

\begin{abstract}
The Seebeck coefficient quantifies the voltage generated across a material in response to a temperature gradient. Recent studies have shown that strong electronic correlations can enhance this coefficient, producing anomalous behavior near half-filling associated with the Mott plateau. This raises the possibility that other interaction scales, not necessarily originating in Mott physics, could give rise to similar enhancements. Here, we investigate the Seebeck coefficient in the presence of attractive interactions, nearest-neighbor interactions, sublattice potentials, and electron-phonon coupling. The Seebeck coefficient is obtained via the Kelvin formula, using entropy data derived from density calculations within determinant quantum Monte Carlo (DQMC). We find that these additional interaction scales can indeed enhance the Seebeck coefficient and further induce multiple sign changes as a function of doping. We show that this anomalous behavior is associated with the opening of a gap in the ground state, as computed via cluster perturbation theory (CPT). Moreover, electron-phonon coupling alone—even in the absence of on-site repulsion—can produce a Seebeck anomaly. We relate these sign changes to a restructuring of the Fermi surface and an accompanying change in its topology, an effect commonly observed in cuprates.
\end{abstract}


\maketitle

\section{Introduction}

The Seebeck coefficient ($S$) quantifies the thermoelectric voltage generated due to a temperature gradient across a material, and is a topic largely scrutinized over the past decades.
In particular, the main goal is to increase the efficiency of the thermoelectric compounds, which is measured by their Figure of Merit,
$ZT = \frac{S^2 \sigma T}{\kappa}$,
with $S$ being the Seebeck coefficient, $\sigma$ the electrical conductivity, $T$ the temperature, and $\kappa$ the thermal conductivity~\cite{rowe2005thermoelectrics,kumar2021thermoelectricity,goldsmid2010introduction}.
It is important to note that optimizing thermoelectric performance is challenging: it requires not only a large $S$, but also high electrical conductivity and low thermal conductivity -- properties that are often difficult to optimize simultaneously.
Typically, materials with high $ZT$ are heavily doped semiconductors or semimetals, in which a delicate balance between the necessary transport properties has to be achieved~\cite{dresselhaus1997prospects,bell2008cooling,snyder2008complex,tomczak2018thermoelectricity}.
Since thermoelectric performance is largely governed by the electronic structure near the Fermi level, strongly correlated materials offer a promising new avenue for its enhancement.
Unlike conventional materials with rigid electronic bands, correlated systems exhibit non-rigid band structures, which may be tuned to achieve more favorable combinations of $S$, $\sigma$, and $\kappa$.

In line with this picture, correlated narrow-gap semiconductors, such as FeSi and FeSb$_{2}$, exhibit exceptionally large Seebeck coefficients that significantly exceed predictions of conventional electronic band structure theories.
For instance, FeSi displays a temperature-dependent $S$ that reaches values above 500 $\mu$V/K at intermediate temperatures~\cite{PhysRevB.90.245146}, while FeSb$_{2}$ is known for exhibiting an even larger low-temperature response to the thermopower, in some cases exceeding $45$ mV/K in absolute value~\cite{Bentien_Steglich_2007}, one of the highest thermopower responses for solids.
There are clear hints that strong electronic correlations play an important role in these materials: they undergo an unusual metal-insulator transition upon heating, with a maximum in the susceptibility and a Curie-Weiss-like behavior at high temperatures~\cite{tomczak2018thermoelectricity}.

Another interesting class of unconventional thermoelectric metals is provided by Na$_x$CoO$_2$\,\cite{Terasaki1997} and K$_x$CoO$_2$\,\cite{Ito2019}. In both compounds, the cobalt ions form triangular CoO$_2$ layers, while the alkali ions (Na or K) are intercalated between them, and control the electron filling.
The electronic states near the Fermi level arise from Co $3d$ orbitals, which are narrow and subject to strong electron-electron interactions.
Although K$_x$CoO$_2$ exhibits only a modest Seebeck coefficient, Na$_x$CoO$_2$ can attain values as high as 100 $\mu$V/K at room temperature, much larger than those found in conventional metals\,\cite{kaurav2009seebeck,lee2006large}.

Cuprates, the paradigm of correlated materials, also exhibit an anomalous Seebeck coefficient.
For instance, in underdoped YBa$_{2}$Cu$_{3}$O$_{6+y}$ (YBCO), a pronounced decrease of $S/T$ to negative values is observed at low temperatures\,\cite{cyr2017anisotropy}.
This behavior indicates that the low-temperature transport is dominated by an electron-like Fermi surface, in stark contrast to the large hole-like Fermi surface expected from Luttinger's theorem in overdoped cuprates.
In other words, the data provide evidence for a Fermi-surface reconstruction (FSR), whose signatures are clearly revealed by thermoelectric measurements.
A similar behavior for the Fermi surface is observed in K$_x$CoO$_2$\,\cite{Ito2019}.

Recent studies have also considered the charge-density wave (CDW) transition in low-dimensional layered materials.
In heterostructures with alternating layers of (PbSe)$_{1 + \delta}$ and VSe$_{2}$, the transition temperature to the CDW phase can be tuned by varying the number of layers.
As discussed in Ref.~\cite{wang2020enhanced}, these layered materials are good candidates for thermoelectric applications: a significant increase of the Seebeck coefficient and the power factor ($PF=S^2\sigma$) was measured in this heterostructure for a single layer of each compound.
This is explained in terms of a low-temperature deviation from the Mott formula for the thermopower due to strong electronic correlations.
Similarly, in the transition-metal dichalcogenide (TMD) TiSe$_{2-x}$, experimental measurements indicate that an enhancement of $S$ occurs alongside a reduction of the CDW order~\cite{Bhatt2014}.
These observations underscore the relevance of CDW physics as a tunable ingredient in designing efficient low-temperature thermoelectric materials.

From a theoretical perspective, the anomaly of the Seebeck coefficient has been studied using effective lattice models.
Early work is reported in Ref.\,\onlinecite{zemljivc2005thermoelectric}, which analyzed the features of $S$ in the one-dimensional Hubbard model by applying linear-response theory combined with exact diagonalization of small chains.
More recently, quantum Monte Carlo (QMC) simulations have been employed in Refs.\,\onlinecite{willdauany2023,roy2024sign,habitzreuter2025specific,wang_quantitative_2023} to study the Seebeck anomaly in the two-dimensional Hubbard model at high temperatures, while Ref.\,\onlinecite{mravlje2022spin} addressed the spin Seebeck effect for the same Hamiltonian, but using finite-temperature Lanczos methods combined with dynamical mean-field theory (DMFT).
For higher dimensions, and using DMFT, Refs.\,\onlinecite{wissgott2010enhancement,wissgott2011effects} investigated how electronic correlations affected the thermoelectric response in Na$_{0.7}$CoO$_2$.

Although the studies in the literature use different methodologies, they generally describe the Seebeck anomaly in terms of the on-site Hubbard model.
However, other interactions may also be relevant.
For instance, in TMDs, electron-electron interactions are relatively weak, while electron–phonon interactions are comparatively strong and can contribute to long-range CDW order, which could affect the Seebeck coefficient as suggested by the experiments mentioned above.
In view of this, the present study aims to explore how interactions beyond the on-site electron-electron (\textit{e-e}) repulsion affect the response of the Seebeck coefficient. In particular, we investigate both local and non-local \textit{e-e} interactions (repulsive or attractive), as well as their interplay with electron-phonon (\textit{el-ph}) coupling.
To this end, we use QMC simulations, exact diagonalization, and cluster approximations to analyze both finite temperatures and ground state properties of different Hamiltonians, employing the Kelvin Formula to obtain the Seebeck coefficient\,\cite{Kelvin2010}.

This paper is organized as follows.
Section~\ref{sec:model_methods} introduces the model Hamiltonians, the determinant quantum Monte Carlo (DQMC) method, and the physical observables considered.
Section~\ref{sec:results} presents results for (A) the attractive Hubbard model, (B) the extended Hubbard model, (C) the Holstein and Hubbard-Holstein models, and (D) the ionic Hubbard model.
In Section~\ref{sec:FSR} we discuss the relation between the Fermi surface reconstruction and the Seebeck anomaly.
Our conclusions and perspectives are given in Section~\ref{sec:conclusions}.

\section{Models and methods}
\label{sec:model_methods}

As we aim to examine the behavior of the Seebeck coefficient on arbitrary interacting systems, we start from a general Hamiltonian
\begin{align}
\label{eq:general_hamiltonian}
\mathcal{H} = \mathcal{H}_{K} + \mathcal{H}_{U} + \mathcal{H}_{V} + \mathcal{H}_{\rm el-ph} + \mathcal{H}_{\rm ph},
\end{align}
in which
\begin{align}
\label{eq:general_hamiltonian2}
\mathcal{H}_{K} &= -t \sum_{\substack{\langle \textbf{i},\textbf{j} \rangle},\sigma} \big( c_{\textbf{i} \sigma}^{\dagger}c_{\textbf{j} \sigma}+ {\rm H.c.} \big) - \sum_{\substack{\textbf{i}}, \sigma} \left(\mu - \epsilon_{\mathbf{i}} \right) n_{\textbf{i},\sigma}~,\\
\mathcal{H}_{U} &= U \sum_{\substack{\textbf{i}}} \big(n_{\textbf{i} \uparrow} - 1/2 \big) \big(n_{\textbf{i} \downarrow} - 1/2\big), \\
\mathcal{H}_{V} &= V  \sum_{\substack{\langle \textbf{i},\textbf{j}\rangle}} n_{\textbf{i}} n_{\textbf{j}}, \\
\mathcal{H}_{\rm el-ph} &= - g \sum_{\substack{\textbf{i}}} n_{\textbf{i}} \hat{X_{\mathbf{i}}}, \\
\mathcal{H}_{\rm ph} &= \frac{1}{2}\sum_{\substack{\textbf{i}}} \left(\hat{P}^{2}_{\mathbf{i}} + \omega^2_{0} \hat{X}^{2}_{\mathbf{i}}\right),
\end{align}
where the sums run over a $L\times L$ square lattice, with $\langle \textbf{i},\textbf{j} \rangle$ denoting nearest neighbors sites, and $\sigma = \{\uparrow, \downarrow\}$ the spin indices.
Here we use the standard second quantization formalism, where the operator $c^{\dagger}_{\mathbf{i} \sigma}$ ($c^{\phantom{\dagger}}_{\mathbf{i} \sigma}$) describes the creation (annihilation) of electrons on a site $\mathbf{i}$, with spin $\sigma$, while $n_{\mathbf{i}\sigma} \equiv c^{\dagger}_{\mathbf{i} \sigma} c_{\mathbf{i} \sigma}$ is the number operator (with $n_{\mathbf{i}} = n_{\mathbf{i}\uparrow} + n_{\mathbf{i}\downarrow}$).
The first term, $\mathcal{H}_{\rm K}$, describes the hopping energy, with H.c.~denoting Hermitian conjugate; it also contains the chemical potential $\mu$, and onsite energies $\epsilon_{\mathbf{i}}$.
The second and third terms, $\mathcal{H}_{\rm U}$ and $\mathcal{H}_{\rm V}$, describe the on-site and the nearest-neighbor interactions with strengths $U$ and $V$, respectively. Such interactions may be defined as positive (repulsive) or negative (attractive).
The fourth term corresponds to local electron-ion interaction, with strength $g$.
Finally, the last term describes optical (Einstein) phonon modes, with $\hat{P}_{\mathbf{i}}$ and $\hat{X}_{\mathbf{i}}$ being momentum and position operators, respectively, and $\omega_{0}$ the frequency. Here, for convenience, we defined the mass of the ions $M$, the Boltzmann ($k_{\rm B}$), Planck ($\hbar$), and lattice ($a$) constants as unity, and set the energy scales in units of $t$.

The thermodynamic and transport properties of the interacting Hamiltonians are obtained by performing Determinant Quantum Monte Carlo (DQMC) simulations~\cite{blankenbecler_1981,hirsch_1983,hirsch_1985,white_1989,santos_introduction_2003,Gubernatis_Kawashima_Werner_2016}.
This is an unbiased methodology, i.e., its inherent approximations can be systematically controlled by increasing computational effort, allowing errors to be reduced arbitrarily close to the exact solution. For instance, for the Hubbard model, we perform the Trotter decomposition for the partition function,
\begin{align}
\nonumber \mathcal{Z} &= \mathrm{Tr}\,
e^{-\beta\mathcal{H}}= \mathrm{Tr}\,
[(e^{-\Delta\tau(\mathcal{H}_{K} + \mathcal{H}_{\rm
U})})^{M}] \\
& \thickapprox \mathrm{Tr}\,
[e^{-\Delta\tau\mathcal{H}_{K}}e^{-\Delta\tau\mathcal{H}_{\rm
U}}e^{-\Delta\tau\mathcal{H}_{K}}e^{-\Delta\tau\mathcal{H}_{\rm
U}}\cdots],
\end{align}
such that $\beta = \Delta \tau M$, where $\Delta \tau$ is the imaginary-time discretization, and $\beta = 1/T$ is the inverse of temperature.
This leads to an error of $\mathcal{O} \left(\Delta \tau\right)^2$, which can be systematically eliminated by performing simulations such that $\Delta \tau \rightarrow 0$. In our results, we ensured $\Delta \tau \leq 0.1$, which is enough to guarantee that the error from the Trotter approximation is smaller than statistical errors from the Monte Carlo sampling procedure.

After the Trotter decomposition, for some interacting systems, we need to perform the Hubbard-Stratonovich transformation. This procedure is applied to transform many-particle (quartic) terms into single-particle (quadratic) ones, i.e.~$e^{-\Delta\tau\mathcal{H}_{\rm U}} \to e^{-\Delta\tau\mathcal{V}}$, at the expense of adding bosonic auxiliary fields $s(\mathbf{i},l)$ coupled to the fermionic ones. In particular, we use discrete real fields, which take values $s(\mathbf{i}, l) = \pm 1$ for each lattice site and time-slice, with $1 \leq l \leq M$.
By performing the fermionic trace, the partition function becomes
\begin{align}\label{eq:partition_func}
{\cal Z} \sim \sum_{\{s(\mathbf{i}, l)\}}
\, \prod_\sigma {\rm det}
\big[ \,I + B_\sigma(M) \cdots B_\sigma(1) \, \big]~,
\end{align}
with $I$ being the identity matrix, $B_\sigma(l) = e^{-\Delta\tau H_{K}} e^{-\Delta\tau V(l)}$, and
$H_{K}$ and $V(l)$ the matrix representations of their corresponding fermionic operators.
The final bosonic trace, $\sum_{\{s(\mathbf{i}, l)\}}$, is performed by standard Monte Carlo techniques, with the product of determinants in Eq.\,\eqref{eq:partition_func} being the statistical weight.
This methodology allows us to calculate the Green's functions and, consequently, any other higher-order physical observables, such as spin, charge, and pair correlation functions.
More details about the DQMC method for the extended Hubbard model can be found in Ref.\,\onlinecite{sousajunior_2024}, while the implementation of the Hubbard-Holstein model is discussed in the Supplementary Information of Ref.\,\onlinecite{costa2020phase}.

However, we should mention that the product of determinants described above might not be positive, resulting in certain configurations having a negative weight in the partition function. This is known as the fermionic minus-sign problem, which we address by a resampling procedure~\cite{hirsch_monte_1982,white_sign_1988,white_numerical_1989,loh_sign_1990}.
That is, when averaging an observable $\mathcal{O}$, one takes the absolute value of the product of the determinant $|p(s)|$ at the cost of keeping track of the average of the sign,
\begin{equation}
\langle \mathcal{O} \rangle = \frac{\sum_s |p(s)| \, \text{sgn}(s) \, \mathcal{O}(s)}{\sum_s |p(s)| \, \text{sgn}(s)} = \frac{\langle \text{sgn} \times \mathcal{O} \rangle}{\langle \text{sgn} \rangle} .
\end{equation}
When the average sign of the determinant is very low, this resampling leads to white noise, increasing the error bars of the statistical sampling.
As this work investigates high-temperature effects, the sign problem is not severe.
Our Monte Carlo simulations are performed with an initial $2 \times 10^3$ Monte Carlo steps for thermalization, followed by $5 \times 10^4$ steps for collecting measurement data.

We are particularly interested in evaluating the Seebeck coefficient, which we obtain from the entropy. The latter is computed (in units of $k_B$) as
\begin{align}
    s(\mu, T) = \int_{\mu_0}^{\mu} d\mu' \left( \frac{\partial n}{\partial T} \right)_{\mu'} \; ,
    \label{eq:entropy_integration}
\end{align}
where
\begin{align}
n(\mu', T) = \frac{\langle \sum_{\mathbf{i}} n_{\mathbf{i}}\rangle}{N}
\label{eq:density}
\end{align}
is the average electronic density at chemical potential $\mu'$ and temperature $T$, with $N=L^2$ the number of lattice sites. The derivative in the integrand is evaluated by a three-point finite difference,
\begin{align}
    \left( \frac{\partial n}{\partial T} \right)_{\mu'}
    & = -\beta^2 \left( \frac{\partial n}{\partial \beta} \right)_{\mu'} \\
    & \approx -\beta^2 \frac{n(\mu',\beta + 2\; \Delta \tau) - n(\mu',\beta - 2\; \Delta \tau)}{4 \; \Delta \tau}~.
\end{align}
We set $\mu_0$ sufficiently negative such that the contribution from $(-\infty,\mu_0)$ to Eq.\,\eqref{eq:entropy_integration} is negligible.

For the Seebeck coefficient, we use the Kelvin formula,
\begin{align}
    S_{\rm Kelvin} &= - \frac{1}{e} \left({\frac{\partial \mu }{\partial T}}\right)_{n} = \frac{1}{e} \left({\frac{\partial s}{\partial n}}\right)_{T}~,
    \label{eq:kelvin_formula}
\end{align}
where $e$ is the electron charge (set to unity hereafter). The second equality follows from Maxwell's relation and has been used in previous studies~\cite{Kelvin2010,Kelvin2013,willdauany2023} to analyze thermoelectric properties of interacting systems (see Ref.~\cite{Kelvin2010} and references therein for comparison between different formulas).
We emphasize that the Kelvin formula is obtained by taking the slow limit ($q_x \rightarrow 0$, followed by $\omega \rightarrow 0$) of the Seebeck coefficient within the Kubo formalism~\cite{Kelvin2010,luttinger1964}. Consequently, it provides a thermodynamic approximation to the thermoelectric properties, with the understanding that it applies primarily in the incoherent transport regime~\cite{Kelvin2013}.
A comparison between such a thermodynamic formula and a full Kubo calculation was performed in Ref.~\onlinecite{wang_quantitative_2023} for the repulsive Hubbard model, also employing DQMC simulations.
Quantitative agreement was observed between the two quantities for the parameters used in our work.
Moreover, Ref.~\onlinecite{wang_quantitative_2023} also shows DQMC results agree with measurements in many materials.

Although comparisons with full Kubo transport calculations have been performed for the repulsive Hubbard model, analogous studies are still lacking for the extended Hubbard, ionic Hubbard, and electron-phonon systems considered here. Therefore, the validity of the Kelvin formula in these cases remains to be established, particularly near gap opening or Fermi-surface reconstruction. Nevertheless, motivated by the results obtained for the repulsive Hubbard model, we extend this thermodynamic perspective to these systems, using the Kelvin formula to investigate their thermopower while acknowledging its limitations as an approximation to the full transport response.

\section{Results}
\label{sec:results}

We organize our results into four limits of the general Hamiltonian in Eq.\,\eqref{eq:general_hamiltonian}. In the next subsection, we consider $U<0$, corresponding to the attractive Hubbard model (AHM). In Subsection~\ref{subsec:ExtHub}, we include nearest-neighbor electron-electron interactions and analyze the extended Hubbard model (EHM) for both attractive and repulsive couplings. We also examine electron-phonon effects by adding local (Holstein) phonon degrees of freedom and studying their interplay with electron-electron interactions within the Holstein-Hubbard model (HHM) in Subsection~\ref{subsec:HHM}. Finally, in Subsection~\ref{subsec:ionic} we discuss the ionic Hubbard model and the emergence of Mott and band-insulating regimes that appear in half and quarter-fillings. In the following, unless otherwise mentioned, we use a temperature of $\beta t = t/T=2$.

\subsection{The attractive Hubbard model}
\label{subsec:attHub}

As discussed in the Introduction, recent studies of the repulsive Hubbard model have shown that electron-electron interactions can enhance the Seebeck coefficient near half-filling~\cite{wang_quantitative_2023,willdauany2023}, accompanied by a doping-dependent sign change in $S_{\rm Kelvin}$. This naturally raises the question of how the thermoelectric response is modified when the on-site interaction is attractive rather than repulsive.
To this end, we consider the Hamiltonian
$$
\mathcal{H} = \mathcal{H}_{K} + \mathcal{H}_{U},
$$
with $U < 0$ and $\epsilon_{\mathbf{i}}=0$, corresponding to the AHM.
At half-filling, the ground state of the AHM supports a supersolid phase, while away from half-filling it exhibits $s$-wave superconducting order~\cite{micnas1990superconductivity}.
In particular, for the interactions and temperatures considered in this work, the system is in the known preformed pairs regime.
Additional details on the model and its phases can be found in Refs~\onlinecite{paiva_attractive_Hubbard_2004,Fontenele_2022,rodrigo_attractive_Hubbard_2024,moreo-scalapino-attractive,monit-attractive}.

\begin{figure}[t]
\includegraphics[scale=0.9]{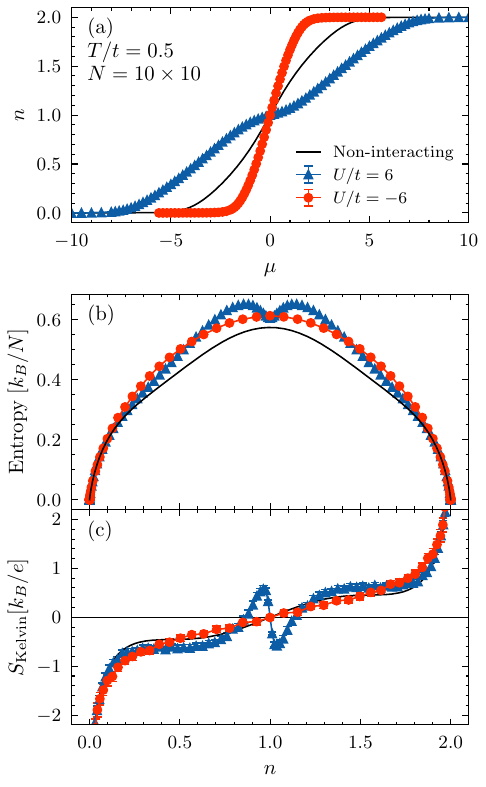}
\caption{(Color online) (a) Density $n$ as a function of chemical potential $\mu$, (b) Entropy, and (c) the Seebeck coefficient as a function of $n$ for non-interacting, attractive, and repulsive interactions $|U/t| = 6$ and temperature $T/t= 0.5$. For derivation and integration, we used $\delta \mu = 0.5$ for $\mu \leq 8$ and $\delta \mu = 0.2$ otherwise.}
\label{fig:dens_entro_attra_repul}
\end{figure}

We begin by examining the electronic density $n(\mu,T)$ as a function of $\mu$ at fixed $T/t=0.5$, displayed in Fig.\,\ref{fig:dens_entro_attra_repul}\,(a). To highlight differences between attractive and repulsive interactions, we consider $U/t=-6$ and $U/t=6$, as well as the noninteracting case.
Notice that, for attractive interactions, the curve indicates a reduced bandwidth compared to the noninteracting case.
This behavior is consistent with the tendency toward double occupancy at strong attractive coupling, which yields a bosonic character and favors occupation of lower-energy states.
In contrast, the repulsive case broadens the spectrum by suppressing double occupancy and pushing spectral weight to higher energies. As expected, there is no Mott plateau at half-filling in the attractive Hubbard model, a feature that can affect the Seebeck response.

We proceed calculating the entropy from $n(\mu)$, which is showed in Fig.\,\ref{fig:dens_entro_attra_repul}\,(b). For $U/t =-6$, the results exhibit a maximum in the entropy at half-filling, similar to the noninteracting case. In contrast, $U/t = 6$ exhibits a minimum as shown previously in the literature~\cite{willdauany2023}, emphasizing that the formation of a Mott insulating state is responsible for such changes in entropy.
We recall that, at the half-filling on any bipartite lattice, the repulsive Hubbard model can be mapped onto the AHM by a partial particle-hole transformation~\cite{arovas2022hubbard}.
Therefore, the entropy response at $n=1$ should be the same for $U/t=6$ or $U/t=-6$. However, slightly away from half-filling, the models exhibit different properties, as shown Fig.\,\ref{fig:dens_entro_attra_repul}.

Using Eq.\,\eqref{eq:kelvin_formula}, we compute the Seebeck coefficient for the attractive, non-interacting, and repulsive Hubbard models, shown in Fig.\,\ref{fig:dens_entro_attra_repul}\,(c).
The models display clear differences.
For the repulsive case, the Seebeck coefficient exhibits a sign change for $n\neq 1$ driven by electron-electron interactions, while in the attractive and non-interacting cases, the only sign change occurs at half-filling due to particle-hole symmetry.
In addition, further away from half-filling, the two models show different behaviors: for example, at fixed $n=0.7$ in Fig.\,\ref{fig:dens_entro_attra_repul}\,(c),  $S_{\rm Kelvin}$ increases in magnitude for $U/t=6$, while in the attractive model it is close to the non-interacting limit.

This latter behavior in the attractive model demands comment. By tuning $U$ in the AHM, one can drive a crossover between Bardeen-Cooper-Schrieffer and Bose-Einstein condensed regimes\,\cite{Fontenele_2022}. In this crossover, a pseudogap phase may appear, characterized by a single-particle excitation gap due to preformed pairs, but without phase coherence. Intuitively, this high-temperature regime can be viewed as a hard-core boson gas, where the singlet pairs are tightly bound and breaking them has a prohibitive energetic cost, which is the nature of the single-particle gap.
Interestingly, for $T/t=0.5$ and $U/t=-6$, the system lies within the parameter range of the pseudogap regime of the attractive model\,\cite{Fontenele_2022}, but no clear signature of this regime is observed in the behavior of $S_{\rm Kelvin}$.
We attribute the lack of dependence in the attractive Hubbard model to its renormalization into a gas of free hard-core bosons, which changes only the effective charge and mass of the particles without affecting their thermoelectric properties. Although lowering the temperature below the critical value may alter this behavior, that regime is not considered here.

\begin{figure}[t]
\includegraphics[scale=0.95]{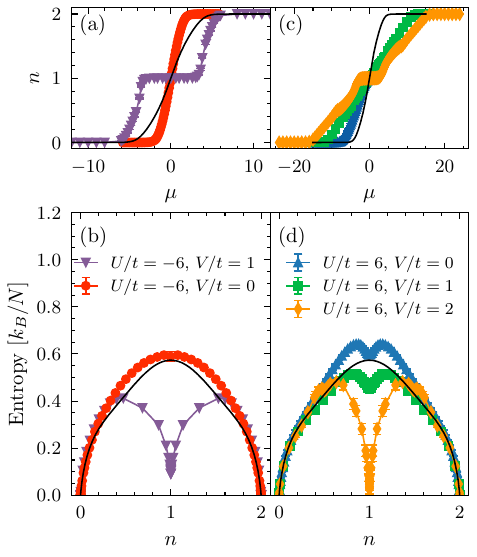}
\caption{(Color online) (a), (c) Density $n$ as a function of chemical potential $\mu$, (b), (d) Entropy as a function of density $n$ for attractive and repulsive interaction $|U/t| = 6$ and temperature $T/t= 0.5$. For $U/t = -6$ and $V/t = 0$ the density profile varies mostly around $\mu = 0$, thus we used $\delta \mu = 0.1$ for $\mu < 2$ and $\delta \mu = 0.2$ otherwise. For $U/t=-6$ and $V/t = 1$, we used $\delta \mu = 0.6$ in the plateaus, where there is a small density variation with $\mu$, and $\delta \mu = 0.3$ otherwise. For the repulsive cases, we used $\delta \mu = 0.2$ for $\mu \leq 8$ and up to $\delta \mu = 1$ for very large chemical potentials.}
\label{fig:dens_entro_attra_exten}
\end{figure}

\subsection{Extended Hubbard model}
\label{subsec:ExtHub}

\subsubsection{Around half-filling}

We now discuss the effects of near-neighbor interactions on the thermoelectric properties. To this end, we examine the extended Hubbard model (EHM)\,\cite{Lin_1995,sousajunior_2024}, which describes fermions on a lattice coupled through one-site ($U$) and first-neighbor ($V$) interactions, whose Hamiltonian reads
$$
\mathcal{H} = \mathcal{H}_{K} + \mathcal{H}_{U} + \mathcal{H}_{V}~,
$$
with $\epsilon_{\mathbf{i}}=0$.
We begin by examining attractive on-site interactions and compare how a nearest-neighbor repulsion $V$ modifies the thermodynamic response around half-filling.

Figure~\ref{fig:dens_entro_attra_exten}\,(a) displays $n(\mu,T)$ for $U/t=-6$ with $V/t=0$ and $V/t=1$. The inclusion of a nearest-neighbor (NN) repulsion opens a gap at half-filling that exceeds the Mott gap for $U/t=6$ in Fig.\,\ref{fig:dens_entro_attra_repul}\,(a).
Intuitively, since a large attractive $|U|$ favors double occupancy, a finite $V>0$ tends to localize the pairs on one sublattice, producing a CDW pattern.
Adding an electron then costs an energy of order $4V$ (on the square lattice), opening a charge gap at half-filling.
We emphasize that, in contrast to the Mott phase in the on-site Hubbard model, which is related to an AFM phase at ground state, the inclusion of NN interactions changes the nature of the charge gap.

\begin{figure}[t]
\includegraphics[scale=0.9]{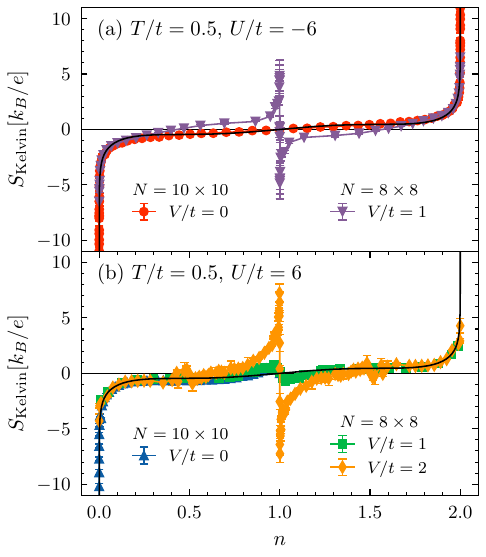}
\caption{(Color online) (a) Seebeck coefficient as a function of density for the attractive case, with $V/t = 0$ and $V/t = 1$ (b) Repulsive case, with $V/t = 0, 1, 2$. Results are for temperature $T/t = 0.5$, with $10 \times 10$ and $8 \times 8$ lattices.}
\label{fig:Seeb_ext_attra_repul}
\end{figure}

Similarly to the previous cases, the opening of the gap produces pronounced changes in the entropy, as shown in Fig.\,\ref{fig:dens_entro_attra_exten}\,(b), including a dip at $n = 1$ for $V/t = 1$. Consequently, the Seebeck coefficient exhibits anomalies near half-filling for these parameters, as shown in Fig.\,\ref{fig:Seeb_ext_attra_repul}\,(a).
Notice that the thermopower anomaly is larger in the case with NN interactions than in the on-site case\,\cite{willdauany2023}, indicating a direct connection between the strength of the anomaly and the size of the Mott gap.
Interestingly, in the CDW phase, the charge gap directly measures the order parameter. Therefore, the anomaly in the Seebeck coefficient can be regarded as an indirect signature of CDW order.

When both on-site and NN interactions are repulsive, we find a similar behavior, but with additional subtleties.
We recall that, at half-filling, the ground state of the extended Hubbard model exhibits a transition from an antiferromagnetic Mott phase to a CDW phase around $U \approx V/4$~\cite{sousajunior_2024}.
Although we are at high temperatures, where long-range antiferromagnetic order is absent, the system retains strong short-range spin or charge correlations, which are enough to open a gap.
Indeed, as shown in Fig.\,\ref{fig:dens_entro_attra_exten}\,(c) for fixed $U/t=6$, increasing $V$ broadens the density plateau at $n = 1$ as a function of the chemical potential.
This behavior is accompanied by a stronger dip in the entropy, shown in Fig.\,\ref{fig:dens_entro_attra_exten}\,(d), and by enhancements of the Seebeck coefficient, as displayed in Fig.\,\ref{fig:Seeb_ext_attra_repul}\,(b).
As the gap increases, the entropy dip deepens, and the Seebeck anomaly becomes more pronounced.

\begin{figure}[t]
    \includegraphics{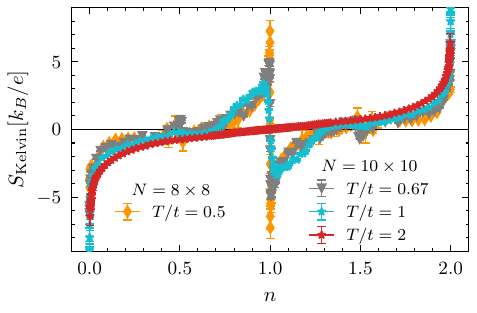}
    \caption{(Color online) Seebeck coefficient as a function of density for different temperatures $T/t = 1, 0.67, 0.5$ for interaction strengths $U/t = 6$ and $V/t = 2$. The lowest temperature was obtained in a simulation on an $8 \times 8$ lattice due to the sign problem. The larger temperatures were simulated with a $10 \times 10$ lattice.}
    \label{fig:Seebeck_n_b2_b15_b1_V2_U6}
\end{figure}

To further illustrate the persistence of the Seebeck anomaly at $V/t=2$, Fig.\,\ref{fig:Seebeck_n_b2_b15_b1_V2_U6} shows the behavior of $S_{\text{Kelvin}}$ at higher temperatures.
Even at $T/t=1$, the anomaly in $S_{\rm Kelvin}$ remains clear.
However, an additional sign change occurs around quarter-filling, whose features are discussed below.

\subsubsection{Around quarter-filling}
\label{sec:extended_ED}

Interestingly, Figs.\,\ref{fig:Seeb_ext_attra_repul} and \ref{fig:Seebeck_n_b2_b15_b1_V2_U6} show an additional sign change of the Seebeck coefficient near quarter-filling, which is robust for $T/t \lesssim 1$.
Indeed, a small plateau is evident in Fig.\,\ref{fig:dens_entro_attra_exten}\,(c) around $n \approx 0.5$, accompanied by small changes in the corresponding entropy in Fig.\,\ref{fig:dens_entro_attra_exten}\,(d).
By contrast, we do not observe this Seebeck anomaly near quarter-filling for $V/t=1$, nor for any cases with $U<0$.
Motivated by the connection between the Seebeck anomaly and the emergence of a Mott gap, we now turn our attention to the sign change in $S_{\rm Kelvin}$ around quarter-filling.

First, we notice that, when $V \gg U$, the single occupation of one of the sublattices becomes strongly favored at quarter-filling for any bipartite lattice.
Under such conditions, a CDW phase should emerge at the ground state~\cite{sousajunior_2024}, which, in turn, would lead to the anomaly of the Seebeck coefficient around $n \approx 0.5$.
To examine this possibility, particularly in the regime $V/t \approx 2$, we perform Exact Diagonalization (ED)~\cite{sandvik_computational_2010} calculations for the extended Hubbard model on a $4\times 4$ lattice with periodic boundary conditions, fixing the particle numbers at $N_{\downarrow}=N_{\uparrow}=4$.
We also analyze the spectral functions at the thermodynamic limit employing Cluster Perturbation Theory (CPT)\,\cite{Hohenadler03, Senechal00, Senechal02, Senechal10} calculations (see Appendix~\ref{appendix:CPT} for details).
The corresponding results are presented below.

\begin{figure}[t]
    \includegraphics[width = \linewidth]{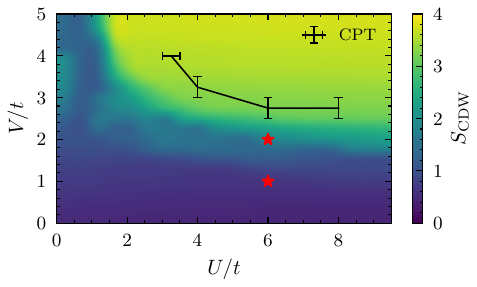}
    \caption{(Color online) Ground state CDW structure factor calculated with ED on a $4 \times 4$ lattice at quarter-filling. The black curve is the phase boundary calculated with CPT. Red stars indicate the parameters of our DQMC simulations for the Seebeck coefficient at quarter-filling.}
    \label{fig:CDW}
\end{figure}

To probe a potential CDW phase, we examine charge-charge correlation functions and their Fourier transform, the charge structure factor, defined as
\begin{equation}
    S_{\text{CDW}} (\bm q) = \frac{1}{N} \sum_{ \mathbf{i} \mathbf{j}} e^{i \bm q \cdot (\mathbf{r}_{i} - \mathbf{r}_{j})} \braket{n_{\bm i} n_{\mathbf{j}}} ~.
\end{equation}
Figure~\ref{fig:CDW} displays the ED results for the maximum of $S_{\text{CDW}}(\mathbf{q})$ as a function of $U/t$ and $V/t$ at quarter-filling, with the magnitude shown as a heat map.
Indeed, there is a region with large charge correlations, with the `transition' curve going asymptotically to both axes.
That is, we find a metallic region (dark blue color) and a CDW region (dark green color) in our ED phase diagram.

To mitigate finite-size effects, we also performed a CPT analysis using $4 \times 4$ clusters. This approach yields the spectral functions (see, e.g., the Appendix), from which the gap opening can be estimated given a numerical threshold. The CPT results are shown as the black line in Fig.\,\ref{fig:CDW}, which are in good agreement with the ED calculations.

To further interpret our DQMC results for the Seebeck coefficient at quarter-filling, we indicate them as red stars in the ground state phase diagram of Fig.\,\ref{fig:CDW}.
Interestingly, for fixed parameters $U/t = 6$ and $V/t = 1$, the system lies in the metallic phase (i.e., no gap). This explains why the Seebeck coefficients at $V/t = 0$ and $V/t = 1$ are quite similar and show no anomalous change of sign at quarter-filling; see Fig.\,\ref{fig:Seeb_ext_attra_repul}\,(b).
In contrast, fixing $U/t = 6$ and $V/t = 2$ places the system at the boundary of a CDW phase, where charge fluctuations are expected.
Although temperature certainly influences the boundaries of these phases, the results provide strong evidence for a close connection between gap formation and the observed Seebeck anomaly.

\subsection{Holstein-Hubbard model}
\label{subsec:HHM}

We now turn to discuss the influence of electron-phonon coupling on the thermopower.
To this end, we analyze the Hubbard-Holstein (HH) model, whose Hamiltonian reads
$$
\mathcal{H} = \mathcal{H}_{K} + \mathcal{H}_{U} + \mathcal{H}_{\rm el-ph} + \mathcal{H}_{\rm ph},
$$
with $\epsilon_{\mathbf{i}} = 0$.
For convenience, we define the dimensionless electron-phonon coupling parameter $\lambda = g^{2}/(\omega^{2}_{0} t)$, which sets the energy scale for polaron formation~\cite{costa2020phase}.
In what follows, we set $\omega_{0} = 1$.

\begin{figure}[t]
\includegraphics[scale=0.9]{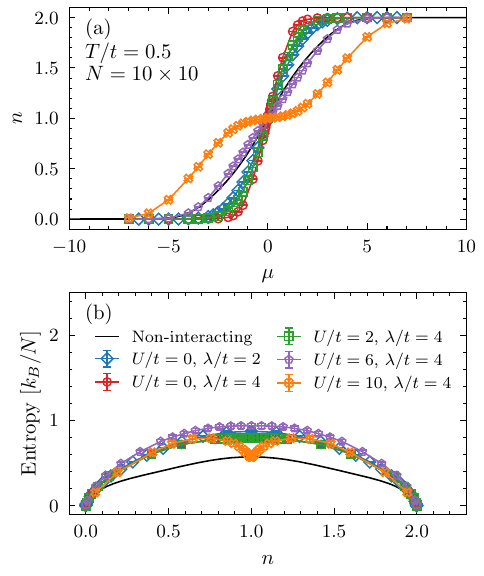}
\caption{(Color online) Density $n$ as a function of chemical potential $\mu$ for different interaction strengths $U/t = 0, 2, 6$ and $10$ and temperature  $T/t= 0.5$, for $\uplambda/t = 2$ and $4$, as a function of density $n$, for the square lattice.}
\label{fig:Dens_entr_hols}
\end{figure}

Figure~\ref{fig:Dens_entr_hols}\,(a) shows the density as a function of chemical potential for several values of $U/t$ and $\lambda/t$.
We emphasize two main features.
First, unlike in the pure Hubbard model, where a Mott plateau is already visible at $U/t=6$ (see Fig.\,\ref{fig:dens_entro_attra_repul}), the presence of electron-phonon coupling smooths out this plateau.
For example, the curve for $U/t = 6$ and $\lambda/t = 2$ is nearly indistinguishable from the noninteracting case.
Second, increasing $\lambda$ at fixed $U = 0$ effectively narrows the bandwidth, leading to a behavior similar to that of the attractive Hubbard model (see Fig.\,\ref{fig:dens_entro_attra_repul}).
These changes induced by the electron-phonon coupling have a strong impact on the entropy, as shown in Fig.\,\ref{fig:Dens_entr_hols}\,(b), where a pronounced dip appears only for the strongest interaction $U/t = 10$.

\begin{figure}[t]
\includegraphics{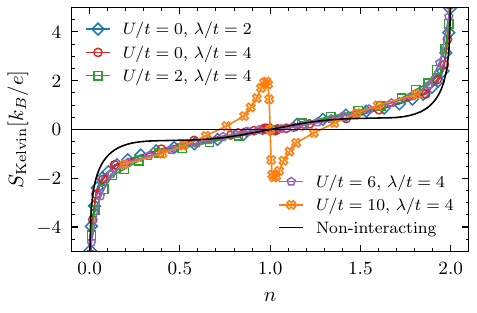}
\caption{(Color online) Seebeck coefficient as a function of density for different interaction strengths $U/t = 0, 2, 6$ and $10$ and temperature  $T/t= 0.5$, for $\uplambda/t = 2$ and $4$. Simulations are for a $10 \times 10$ square lattice.}
\label{fig:Seeb_Hols}
\end{figure}

Figure~\ref{fig:Seeb_Hols} shows $S_{\text{kelvin}}$ as a function of $n$ for the HH model, using the same parameters as in Fig.\,\ref{fig:Dens_entr_hols}.
The Seebeck anomaly appears only at large $U$, where the system enters the Mott insulating phase.
This behavior is expected: the electron-phonon coupling generates retarded attractive interactions that effectively reduce the repulsive Hubbard $U$.
As a result, increasing $\lambda$ weakens the Seebeck anomaly relative to the pure Hubbard model.
In this sense, the presence of both electron-electron and electron-phonon interactions suppresses the anomaly of the Seebeck coefficient.

\begin{figure}
    \centering
    \includegraphics[width=\linewidth]{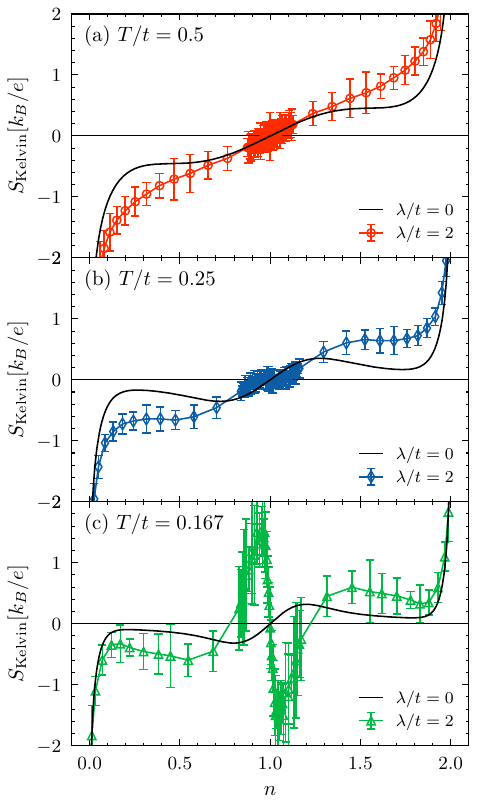}
    \caption{(Color online) Seebeck coefficient for $U = 0$ at temperatures (a) $T/t = 0.5$, (b) $T/t = 0.25$ and (c) $T/t = 0.167$. Closed symbols for the $10 \times 10$ lattice with $\lambda/t = 2$. The black continuous line is calculated with $\lambda = 0$ in the thermodynamic limit for each temperature. The gap is small in terms of $\mu$, requiring many density values to resolve the behavior of $S_{\rm Kelvin}$ around half-filling. The $\mu$-grid is $\delta \mu = 0.01$ for $\mu \leq 2$ and $\delta \mu = 0.2$ otherwise.}
    \label{fig:Seeb_Hols_temp}
\end{figure}

This result may appear somewhat counterintuitive, since electron-phonon systems can develop a CDW insulating phase, which may in turn enhance the Seebeck coefficient.
To clarify the effects of phonons on $S_{\rm Kelvin}$ response, we analyze the pure Holstein model (i.e., setting $U=0$).
Figure \ref{fig:Seeb_Hols_temp} shows the temperature dependence of $S_{\rm Kelvin}$ at fixed $\lambda/t = 2$.
For this coupling strength, the Holstein model undergoes a finite-temperature transition into a CDW phase at $T_{\rm CDW} = 0.185t$~\cite{natanael_phonons_2018,batrouni_phonons_transition_2019}.
Indeed, the Seebeck anomaly only becomes visible below this energy scale, as illustrated in Fig.\,\ref{fig:Seeb_Hols_temp}\,(c) for $T = 0.167t$.
This behavior reflects the fact that, although strong charge fluctuations exist above $T_{\rm CDW}$, a full gap forms only once long-range order is established.
Consequently, as mentioned before, the onset of anomalies in the Seebeck coefficient serves as an indirect signature of CDW long-range order.

A final comment is in order.
When strong attractive interactions (either direct or mediated) are present, a pseudogap phase may emerge.
This regime is characterized by tightly bound local electron pairs, similar to a BEC.
As a result, varying the chemical potential mainly adds or removes pairs of electrons, producing a stair of gaps that become nearly imperceptible at finite temperature~\cite{xiao2021charge}.
Then, $n(\mu)$ exhibits a smooth behavior, hiding the presence of an actual gap (or pseudogap). In such situations, extracting $S_{\rm Kelvin}$ from thermodynamic quantities may become ill-defined.

\subsection{Ionic Hubbard model}
\label{subsec:ionic}

The last scenario we examine is the one in which both band and Mott insulating phases may arise. This case can be investigated using the ionic Hubbard model\,\cite{bouadim_metallic_2007}, whose Hamiltonian reads
$$
\mathcal{H} = \mathcal{H}_{K} + \mathcal{H}_{U},
$$
with $U > 0$ and a staggered on-site potential $\epsilon_{\mathbf{i}} = \pm \Delta$ (positive on one sublattice and negative on the other).
The presence of a staggered potential modifies the correlations at half-filling, leading to a competition between band and Mott insulating phases, as discussed, e.g., in Ref.\,\onlinecite{bouadim_metallic_2007}.
Similar changes in the correlation effects are also observed at quarter-filling. Therefore, we analyze these two cases separately.

\subsubsection{Around half-filling}

\begin{figure}
    \centering
    \includegraphics{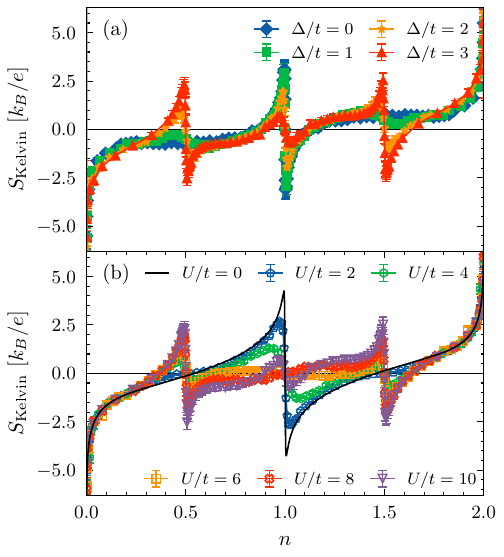}
    \caption{(Color online) Seebeck coefficient of the ionic Hubbard model at $T/t = 0.5$ with (a) fixed $U/t = 10$ and (b) fixed $\Delta/t = 3$.}
    \label{fig:ionic_seebeck}
\end{figure}

As in the previous cases, we analyzed the density profiles, local density of states, and entropy. Since the connection between density plateaus and entropy peaks was already highlighted in the previous Hamiltonians, here we show only the Seebeck coefficient in Figure~\ref{fig:ionic_seebeck}.

We first analyze the half-filling regime.
For $\Delta = 0$, the system is a Mott insulator at strong $U$.
As the staggered potential $\Delta$ increases for fixed $U$, the system first goes from a Mott insulator to a metal and then enters a band-insulator phase~\cite{bouadim_metallic_2007}.
Panel (a) of Figure~\ref{fig:ionic_seebeck} shows the Seebeck coefficient for a fixed on-site interaction $U/t = 10$ and temperature $T/t = 0.5$ for a few intermediate values of $\Delta$. As $\Delta$ grows, we detect a decrease in the Seebeck response around $n = 1$, indicating a transition to the metallic state where there is no Seebeck anomaly.

On the other hand, for $U=0$ and $\Delta > 0$, the system exhibits two bands, being a band insulator at half-filling, with a gap of approximately $2\Delta$. This behavior is observed in the Seebeck coefficient for $\Delta/t = 3$ and $U=0$, shown in Fig.\,\ref{fig:ionic_seebeck}\,(b), as a black solid curve. The abrupt change in $S_{\rm Kelvin}$ arises from the band occupations, while the behavior within each band ($0 < n < 1$ and $1 < n < 2$) remains quite similar to that of the single-band case shown in Figs.\,\ref{fig:Seeb_ext_attra_repul} and \ref{fig:Seeb_Hols}. As $U$ increases, however, this abrupt variation in $S_{\rm Kelvin}$ is suppressed due to the emergence of metallic behavior at half-filling. For still larger $U$, the Seebeck response around $n=1$ increases again as the Mott insulating phase develops. Therefore, the Seebeck coefficient clearly distinguishes the metallic, band-insulating, and Mott-insulating regimes, capturing these transitions.

\subsubsection{Around quarter-filling}

We now turn our attention to the quarter-filling regime, in which the ground state is less clear.
Figure \ref{fig:ionic_seebeck}\,(a) shows that, as $\Delta$ increases for fixed $U/t=10$,  the Seebeck response for $n \lesssim  0.5$ has a change from a value close to the noninteracting case (for $\Delta=0$) to a large response with opposite sign (for $\Delta/t =2$ and 3).
Indeed, a clear anomaly in the Seebeck is observed around $n=0.5$.

The anomaly in $S_{\rm Kelvin}$ becomes more visible when fixing $\Delta$ and varying $U$, as presented in Figure~\ref{fig:ionic_seebeck}\,(b).
As before, we emphasize that the zero-crossing of the Seebeck for $U=0$ does not indicate a Seebeck anomaly, resulting purely from the lower $-\Delta$ band being half-filled.
In this sense, it is a trivial zero, in the same way that $n = 1$ leads to $S_{\text{Kelvin}} = 0$ due to particle-hole symmetry.
However, this scenario changes as the interaction $U$ increases: we notice a non-trivial $S_{\text{Kelvin}} = 0$, signaled by steep increases in value and additional sign changes.
This suggests a transition from a metallic state at low $U$ to a Mott insulating phase at high $U$, which corresponds to a density plateau at $n = 0.5$ in this model (not shown).
In particular, note the significant increase in the Seebeck response around $n = 0.5$ for $U/t = 10$ in comparison with the Extended Hubbard model for the same filling, shown in Fig.~\ref{fig:Seeb_ext_attra_repul}.

Intuitively, one may expect the ground state of the ionic Hubbard model at quarter-filling to be an antiferromagnetic Mott insulator. This expectation can be motivated by considering the limit $\Delta \gg t$. In this regime, one sublattice is completely filled, and hopping to the other sublattice occurs only virtually. When $U$ is introduced, these virtual hopping processes generate an exchange coupling between the nearly localized electrons on the first sublattice, thereby giving rise to antiferromagnetic correlations.

\begin{figure}
    \centering
    \includegraphics{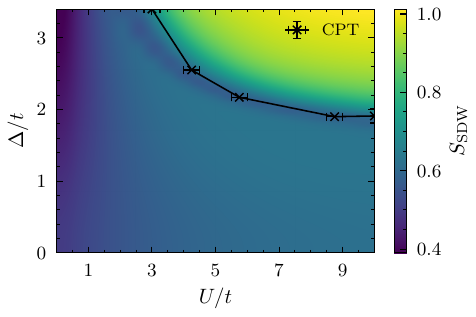}
    \caption{(Color online) SDW structure factor for the ionic Hubbard model calculated from ED. Black crosses show where the gap opens, as calculated from CPT.}
    \label{fig:ionic_SDW}
\end{figure}

To probe antiferromagnetic correlations at the ground state, we perform ED calculations for the ionic Hubbard model and compute the SDW structure factor, defined as
\begin{equation}
    S_{\text{SDW}} (\bm q) = \frac{1}{N} \sum_{ \mathbf{i} \mathbf{j}} e^{i \bm q \cdot (\mathbf{r}_{i} - \mathbf{r}_{j})} \braket{\left( n_{\mathbf{i}\uparrow} - n_{\mathbf{i}\downarrow} \right) \left( n_{\mathbf{j}\uparrow} - n_{\mathbf{j}\downarrow} \right)}.
\end{equation}
Fig.\,\ref{fig:ionic_SDW} shows the maximum of $S_{\text{SDW}}$ at quarter-filling as a heat map, as function of $U/t$ and $\Delta/t$.
For small $U/t$ or $\Delta/t$, antiferromagnetic correlations remain weak, and the system stays metallic.
This explains why no anomaly is observed in the Seebeck coefficient for $\Delta/t = 0$ or 1, even at $U/t = 10$, as shown in Fig.\,\ref{fig:ionic_seebeck}\,(a).
A similar conclusion applies for small values of $U/t$ at fixed $\Delta/t=3$, as illustrated in Fig.\,\ref{fig:ionic_seebeck}\,(b).
For larger values of $U/$ and $\Delta/t$, there is an insulating phase, with strong antiferromagnetic correlations, as displayed in Figure~\ref{fig:ionic_SDW}.

Repeating the same procedure outlined before for the extended Hubbard model, we employ CPT to estimate the phase transition into the Mott state (see Appendix~\ref{appendix:CPT}).
Indeed, there is good agreement between the gap opening, the increase in the SDW structure factor, and the anomaly of the Seebeck coefficient around quarter-filling in Fig.~\ref{fig:ionic_seebeck}.

\section{Fermi surface reconstruction}
\label{sec:FSR}

The sign change in the Seebeck coefficient is a hallmark of a change in the type of carrier, either electron-like or hole-like.
Therefore, in our case, this behavior may be associated with a Fermi surface reconstruction driven by electronic interactions \cite{paiva_osborne,Sakai, Civelli,Helena}.
To this end, we need to examine the spectral functions $A(\mathbf{k},\omega)$, as they provide direct insight into the redistribution of spectral weight and the emergence of new quasiparticle states.
We calculate the spectral functions by performing an analytical continuation for the imaginary-time-dependent Green's functions,
\begin{align}
G(\mathbf{k},\tau) = \int_{-\infty}^{\infty} \mathrm{d}\omega\, A(\mathbf{k},\omega) \,
\frac{e^{-\omega \tau}}{e^{-\beta \omega} + 1}~.
\label{eq:aw}
\end{align}
This inversion can be done with the Maximum Entropy method~\cite{jarrell96,sandvik_stochastic_1998}.

\begin{figure}[t]
\centering
\includegraphics[scale=1]{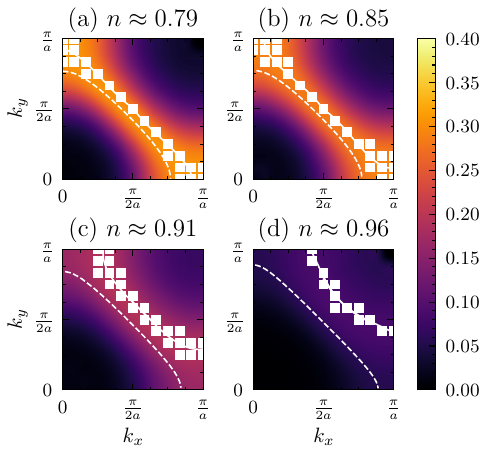}
\caption{The spectral function at the Fermi level $A(\bm k, \omega = 0)$ (in arbitrary units) for $U/t = 10$, $T/t = 0.5$, and average densities (a) $n \approx 0.79$, (b) $n \approx 0.85$, (c) $n \approx 0.91$ and (d) $n \approx 0.96$. The dashed line is the exact calculation for the noninteracting case in the thermodynamic limit, while the squares mark the highest values of $A(\bm k, \omega = 0)$ simulated in a $24 \times 24$ lattice. The white continuous line is a fit through these highest points. The upper panels display data for density values such that $S_{\text{Kelvin}} < 0$, while the lower panels display data for $S_{\text{Kelvin}} > 0$. The change of sign of the Seebeck coefficient happens at $n = 0.87 \pm 0.01$.
}
\label{fig:FS_U10}
\end{figure}

To facilitate the discussion, we focus on the on-site Hubbard model, although the implications of our analysis are general.
We examine the spectral function $A(\mathbf{k}, \omega = 0)$ at the Fermi level for $U/t = 10$ and different electronic densities, as shown in Fig.\,\ref{fig:FS_U10}. Calculations were performed on a $24 \times 24$ lattice, and the data were interpolated for plotting; white square symbols mark the peaks of $A(\mathbf{k}, \omega = 0)$, while the dashed white line indicates the corresponding noninteracting Fermi surface (FS).
The upper panels of Fig.\,\ref{fig:FS_U10} display $A(\mathbf{k}, \omega = 0)$ for densities where $S_{\rm Kelvin} < 0$. In these cases, the spectral peaks follow the noninteracting contour, revealing an electron-like closed FS.
In contrast, for densities where the Seebeck coefficient changes sign, the situation is quite different, as shown in the lower panels of Fig.\,\ref{fig:FS_U10}.
Here, the peaks of $A(\mathbf{k}, \omega = 0)$ shift in a manner consistent with a hole-like FS.
Note that, while the shape of the FS follows the sign of $S_{\rm Kelvin}$, the density resolution of our simulation does not allow us to claim this change of behavior happens exactly at the same density.
This behavior persists for other values of $U$ and temperature (not shown), directly associating the sign change in the Seebeck coefficient with an FSR.

\begin{figure}
    \centering
    \includegraphics[width=\linewidth]{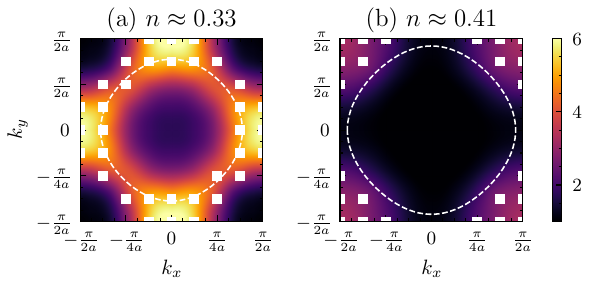}
    \caption{
    The spectral function at the Fermi level $A(\bm k, \omega = 0)$ (in arbitrary units) for $U/t = 10$, $T/t = 0.5$, $\Delta/t = 3.0$ and average densities (a) $n \approx 0.3$, and (b) $n \approx 0.41$. The dashed line is the exact calculation for the noninteracting case ($U/t = 0$) in the thermodynamic limit, while the squares mark the highest values of $A(\bm k, \omega = 0)$ simulated in a $16 \times 16$ lattice.  The left panel displays data for a density value such that $S_{\text{Kelvin}} < 0$, while the lower panel data is for $S_{\text{Kelvin}} > 0$.
    }
    \label{fig:FS_U10_delta3}
\end{figure}

To further clarify the connection between FSR and the Seebeck anomaly, we now consider the ionic Hubbard model, focusing on the regime near quarter-filling. We choose the case with the strongest interactions, $U/t = 10$ and $\Delta/t = 3$, for which the Seebeck coefficient changes sign around $n = 0.36$. We therefore analyze the densities $n \approx 0.33$ and $n \approx 0.41$, whose corresponding Fermi surfaces are shown in Figs.~\ref{fig:FS_U10_delta3}\,(a) and (b), respectively. Notice that, for $n \approx 0.33$, the Seebeck coefficient behaves similarly to the noninteracting case, as shown in Fig.\,\ref{fig:ionic_seebeck}, while the Fermi surface exhibits enhanced spectral weight around the noninteracting contour. By contrast, for $n \approx 0.41$ [Fig.~\ref{fig:FS_U10_delta3}\,(b)], the spectral weight shifts toward the edges near $\mathbf{k} = (\pi/4,\pi/4)$, indicating the occurrence of an FSR.

\begin{figure}
    \centering
    \includegraphics[width=\linewidth]{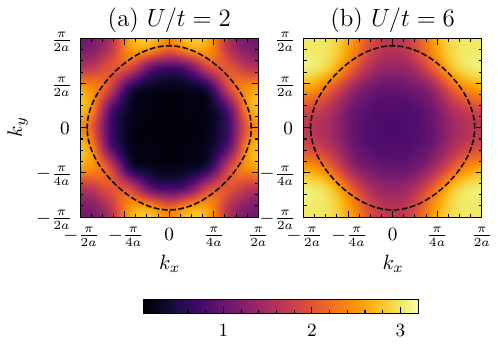}
    \caption{
    The spectral function at the Fermi level $A(\bm k, \omega = 0)$ (in arbitrary units) for $T/t = 0.5$, $\Delta/t = 3.0$, $n \approx 0.41$ and interactions (a) $U/t = 2$, (b) $U/t = 6$, and $U/t = 10$. The dashed line is the exact calculation for the noninteracting case ($U/t = 0$) in the thermodynamic limit, while the squares mark the highest values of $A(\bm k, \omega = 0)$ simulated in a $16 \times 16$ lattice.
    }
    \label{fig:FS_n0.41_delta3}
\end{figure}

A similar evolution of the spectral weight is observed at fixed density $n \approx 0.41$ and $\Delta/t = 3$ as the interaction strength $U$ is varied, as shown in Fig.\,\ref{fig:FS_n0.41_delta3} for (a) $U/t = 2$ and (b) $U/t = 6$. The corresponding Seebeck coefficients are presented in Fig.\,\ref{fig:ionic_seebeck}\,(b). For weak interaction, where $S_{\rm Kelvin}$ remains similar to the noninteracting case, the spectral weight approximately follows the Fermi surface of the $U=0$ system. In contrast, for $U/t = 6$, where the Seebeck coefficient changes sign, the spectral weight is clearly shifted toward the corners of the Brillouin zone.

In both cases, whether upon varying $n$ or $U$, we observe a direct connection between the sign change of the Seebeck coefficient and the occurrence of an FSR, highlighting the usefulness of this transport coefficient as a probe of Fermi-surface topology. Consistent with this interpretation, the FSR is also closely related to the sign of the Hall coefficient $R_H$, as discussed in Ref.\,\onlinecite{wang2020dc}.

\section{Conclusions}
\label{sec:conclusions}

In this work, we investigated the thermopower of interacting two-dimensional lattice systems by means of determinant quantum Monte Carlo simulations, complemented by exact diagonalization and cluster perturbation theory. Focusing on the entropy and on the Seebeck coefficient obtained from the Kelvin formula, we analyzed how the thermoelectric response evolves with filling, temperature, and interaction strength across several effective Hamiltonians, namely the attractive Hubbard model, the extended Hubbard model, the Holstein and Hubbard-Holstein models, and the ionic Hubbard model.

Our results show that strong interactions, together with sufficiently low temperatures, can drive anomalous sign changes in the Seebeck coefficient. While in the Holstein-Hubbard model this behavior remains confined to the vicinity of half-filling, as in the conventional repulsive Hubbard model, the extended Hubbard and ionic Hubbard models also display anomalous sign reversals near quarter-filling. In these cases, the comparison with structure factors from exact diagonalization and gap estimates from cluster perturbation theory indicates that the onset of the Seebeck anomaly is closely connected to the emergence of insulating behavior and to the corresponding metal-insulator boundaries. In addition, we showed that, in the pure Holstein model, the appearance of the Seebeck anomaly tracks the onset of CDW order even in the absence of on-site repulsion.

Our results also reveal that additional interaction scales can significantly enhance the thermopower response. In particular, nearest-neighbor interactions strongly amplify the Seebeck anomaly near half-filling in the extended Hubbard model, while the ionic Hubbard model displays the largest response near quarter-filling among the parameter sets examined here. By contrast, the attractive Hubbard model does not exhibit an anomalous sign change on its own, requiring the inclusion of repulsive nearest-neighbor interactions to produce such a feature at half-filling.

Finally, by analyzing the spectral weight at the Fermi level, we found that the changes in sign of the Seebeck coefficient are accompanied by a reconstruction of the Fermi surface (or, at least, a clear change in the spectral weight), with clear deviations from the noninteracting topology across the transition. Taken together, these results establish the Seebeck coefficient as a sensitive probe of interaction-driven changes in electronic structure, including Mott gap formation and Fermi-surface reconstruction.
An important direction for future research is the investigation of the spin Seebeck effect, whereby a temperature gradient induces a spin current. Recent experiments have reported nontrivial behavior of the spin Seebeck coefficient in correlated antiferromagnets\,\cite{spin_seebeck_experiment_2025}. These findings suggest that the methodology developed here may be naturally extended to the study of spin thermoelectric response in strongly correlated systems. Phonon heat transport effects could also be considered in the future.

\section*{Acknowledgements}
We thank R.G.~Pereira for enlightening discussions about this work.
The authors acknowledge financial support from the Brazilian funding agencies Conselho Nacional de Desenvolvimento Cient\'\i fico e Tecnol\'ogico (CNPq), Coordena\c c\~ao de Aperfei\c coamento de Pessoal de Ensino Superior (CAPES), and Fundação de Amparo \`a Pesquisa do Estado do Rio de Janeiro (FAPERJ).
T.P.~acknowledges support from FAPERJ Grants No.~E-26/210.100/2023, E-26/210.781/2025, and   E-26/200.230/2026, and CNPq grant Nos.~308335/2019-8 and 442072/2023-6.
N.C.C.~acknowledges support from FAPERJ Grants No.~E-26/200.258/2023 [SEI-260003/000623/2023] and E-26/210.592/2025 [SEI-260003/004500/2025], CNPq Grant No.~308130/2025-1, Serrapilheira Institute Grant No.~R-2502-52037, and Alexander von Humboldt Foundation.
M.A.H. thanks CAPES for funding, the International Institute of Physics of UFRN (IIP-UFRN), where part of this work was completed (Grant No. 1699/24 IIF-FINEP), and CENAPAD-SP (National Center for High Performance Processing in São Paulo) for computational resources, project UNICAMP/FINEP-MCTI.

\appendix

\section{Cluster Perturbation Theory}
\label{appendix:CPT}

\begin{figure}
    \centering
    \includegraphics[]{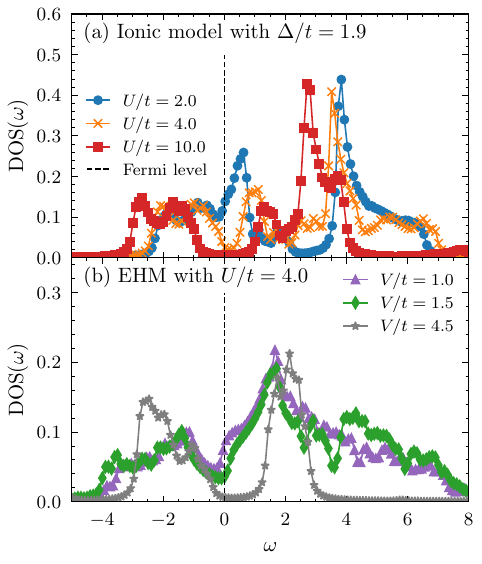}
    \caption{Density of states for (a) the ionic Hubbard model at fixed $\Delta/t = 1.9$ and various values of $U$ in the $4 \times 2$ cluster and (b) the EHM at fixed $U/t = 4$ and various values of $V/t$ in the $4 \times 3$ cluster. The Fermi level is denoted by a vertical dashed line.}
    \label{fig:DOS}
\end{figure}

To evaluate the $T=0$ gap, we employed Cluster Perturbation Theory (CPT), a convenient method to extract spectral weights and the electronic density of states (DOS) of correlated systems.
In brief, CPT tiles the infinite lattice into finite-size clusters, where the Hamiltonian is solved exactly, while the coupling between clusters is treated perturbatively.
Pedagogical introductions to this method can be found in Refs.~\cite{Hohenadler03, Senechal00, Senechal02, Senechal10}.

Within this framework, the Hamiltonian is decomposed as
\begin{align}
    H = H_{\text{cluster}} + T~,
    \label{eq:Hcpt}
\end{align}
where $H_{\text{cluster}}= K + H_{\text{int}}$ denotes the intra-cluster Hamiltonian.
Here, $K = \sum_{\alpha, \beta} t_{\alpha,\beta}c_{\alpha}^{\dagger}c_{\beta}$ and $H_{\text{int}} = -U  \sum_{\textbf{i}}  \left(n_{\textbf{i},\uparrow} -1/2 \right)\left(n_{\textbf{i},\downarrow} -1/2 \right) + \Delta\sum_{i}(-1)^{i}n_{i}$ being the hopping and the local terms of the Hamiltonian \eqref{eq:general_hamiltonian}, respectively.
In the term $K$, $t_{\alpha,\beta}$ are the hopping amplitudes between sites located inside a given cluster.
The term $T = \sum_{\alpha, \beta}V_{\alpha, \beta}c_{\alpha}^{\dagger}c_{\beta} $ is the one that connects different clusters of the superlattice (i.e., intercluster sites) where, $V_{\alpha,\beta}$ represents the hopping amplitudes between sites located in distinct clusters.
Thus, the Green’s functions are then related by
\begin{align}
    \textbf{G}(\tilde{\omega})^{-1} = \mathbf{G}'(\tilde{\omega})^{-1} - \mathbf{V},
    \label{eq:TFGFcpt1}
\end{align}
where $\tilde{\omega}=\omega-\mu$ is the shifted frequency, $\mathbf{G}(\tilde{\omega})$ is the Green’s function of the infinite lattice, and $\mathbf{G}'(\tilde{\omega})$ is the exact Green’s function of the finite cluster.

Since the cluster superlattice has a reduced Brillouin zone (RBZ), any wavevector ${\bf k}$ in the original Brillouin zone can be mapped onto a ${\bf \tilde{k}}$ in the RBZ via ${\bf k}={\bf \tilde{k}}+{\bf K}$, with ${\bf K}$ an element of the reciprocal superlattice.
In the RBZ, Eq.\,\eqref{eq:TFGFcpt1} becomes
\begin{align}
    \textbf{G}(\textbf{\~k},\tilde{\omega})^{-1} = \mathbf{G}'(\tilde{\omega})^{-1} - \mathbf{V}(\textbf{\~k})~,
    \label{eq:TFGFcpt2}
\end{align}
which eventually leads to the CPT Green's functions
\begin{align}
    G_{\text{CPT}}(\mathbf{k},\tilde{\omega}) = \frac{1}{N_{c}}\sum_{a,b}e^{-i\mathbf{k}(\mathbf{r}_{a} - \mathbf{r}_{b})}G_{a,b}(\mathbf{\tilde{k}},\tilde{\omega})~,
    \label{eq:GreenCPT}
\end{align}
where $N_{c}$ is the number of sites in the cluster and $\mathbf{r}_{a(b)}$ are the positions of the intra-cluster sites.

The CPT Green's functions take into account the leading order in the interaction term, and allow one to obtain the spectral weight
\begin{align}
    A_{0}(\mathbf{k},\tilde{\omega}) = -\frac{1}{2\pi}\text{Im}[G_{\text{CPT}}(\mathbf{k},\tilde{\omega})],
    \label{eq:A0}
\end{align}
and the DOS
\begin{align}
    N(\tilde{\omega}) = \frac{1}{N}\sum_{\mathbf{k}} A_{0}(\mathbf{k},\tilde{\omega}).
    \label{eq:DOS}
\end{align}

We analyze the behavior of the DOS at the Fermi level for $n = 0.5$ to identify the transition to a metal–magnetic-insulator phase.
Fig.~\ref{fig:DOS} shows the DOS as a function of $\omega$ for two models, with the Fermi level denoted by a dashed line. Panel (a) is the ionic model with a fixed sublattice potential $\Delta/t = 1.9$ and various values of $U/t$. For $U/t = 2$, we detect a gap at $\omega = 0$. As the interaction is increased, the gap is reduced. For a large interaction $U/t = 10$, no gap is detected.
Meanwhile, Fig.~\ref{fig:DOS} (b) shows the EHM for a fixed $U/t = 4.0$ and various values of $V/t$. For a weak neighbor repulsion $V/t = 1.0$, there is a gap, while for a larger $V/t = 4.5$ the gap has closed.

\bibliography{references}

\end{document}